\documentclass[12pt]{article}

\usepackage{newtxtext,newtxmath}

\usepackage[letterpaper,margin=1in]{geometry}

\usepackage{floatrow}
\usepackage[label font=bf,labelformat=simple]{subfig}
\renewcommand{\thesubfigure}{\sffamily\bfseries\Alph{subfigure}}
\usepackage{graphicx}
\usepackage{multirow}%
\usepackage{setspace} 
\usepackage{xcolor}%
\usepackage{textcomp}%
\usepackage{manyfoot}%
\usepackage{booktabs}%
\usepackage{algorithm}%
\usepackage{algorithmicx}%
\usepackage{algpseudocode}%
\usepackage{listings}%
\usepackage{amsmath}
\usepackage{makecell}
\usepackage{media9}
\usepackage{hyperref}
\usepackage{geometry}
\usepackage{physics}
\usepackage{soul}
\renewenvironment{abstract}
	{\quotation}
	{\endquotation}

\date{}

\renewcommand\refname{References and Notes}

\makeatletter
\renewcommand{\fnum@figure}{\textbf{Figure \thefigure}}
\renewcommand{\fnum@table}{\textbf{Table \thetable}}
\makeatother

\usepackage{scicite}

\usepackage{url}

\def\scititle{Extreme Solar Storm Reveals Causal Interactions in Space Weather}
\title{\bfseries \boldmath \scititle}

\author{
	Xinan~Dai$^{1}$,
	Haiyang~Fu$^{1\ast}$,
	Zichong~Yan$^{1}$,
    Zitong~Wang$^{1}$,
    Feng~Xu$^{1\ast}$,\and
    Chi~Wang$^{2}$,
    Yuhong~Liu$^{1}$,
    Ya-Qiu~Jin$^{1}$\and
	\small$^{1}$Key Laboratory for Information Science of Electromagnetic Waves (MoE),\and\small Fudan University, Shanghai \& 200433, China.\and
	\small$^{2}$State Key Laboratory of Space Weather, National Space Science Center,\and\small Chinese Academy of Sciences, Beijing \& 100190, China.\and
	\small$^\ast$Corresponding author. Email: haiyang\_fu@fudan.edu.cn; fengxu@fudan.edu.cn\and 
}

\begin{document} 

\maketitle

\begin{abstract} \bfseries\boldmath\singlespacing
Solar storms perturb Earth's magnetosphere, triggering geomagnetic storms that threaten space-based systems and infrastructure. Despite advances in spaceborne and ground-based observations, the causal chain driving solar-magnetosphere-ionosphere dynamics remains elusive due to multiphysics coupling, nonlinearity, and cross-scale complexity. This study presents an information-theoretic framework to decipher interaction mechanisms in extreme solar geomagnetic storms across intensity levels within space weather causal chains, using 1980-2024 datasets. Unexpectedly, we uncover auroral spatial causality patterns associated with space weather threats in the Arctic during May 2024 extreme storms. By integrating causal consistency constraints into spatiotemporal modeling, SolarAurora outperforms existing frameworks, achieving superior accuracy in forecasting May/October 2024 events. These results advance understanding of space weather dynamics and establish a promising framework for scientific discovery and forecasting extreme space weather events.
\end{abstract}

Solar storms, driven by the variable activity of the Sun, are significant space weather phenomena capable of disrupting Earth's technological systems and natural environment \cite{MINNIS1958,Normile2000,Voosen2018,Greshko2024}. These storms manifest primarily as solar flares and coronal mass ejections (CMEs), both of which occur in the outer corona of the Sun \cite{Gou2019,Veronig2020,Kusano2020}. When CMEs and flare-associated electromagnetic emissions interact with near-Earth environment, they can induce geomagnetic storms, which may lead to substantial disturbances in Earth's magnetic field and a range of space weather effects \cite{SOICHER1976,Walsh2014,Keiling2019,Milan2017a}. The ionosphere, a critical region of the atmosphere of the Earth, plays a key role in propagating the effects of solar storms \cite{Lyon2000,Kelley2012}. Fluctuations of ionospheric electron density can degrade high-frequency communication signals and disrupt global navigation positioning systems \cite{Voosen2017,Gibney2017}. With society’s increasing dependence on satellite and ground-based infrastructures \cite{Smith2024,Wang2010,Yue2024}, these systems face unprecedented risks from solar storm disruptions, 
therefore accurate prediction and mitigation strategies \cite{Baker2002, Hapgood2012} are essential for safeguarding technological infrastructure and enhancing human preparedness against space weather anomalies \cite{Bolduc2002,Castelvecchi2024}. Although space weather has been studied for over a century \cite{Chapman1940,Hasegawa2004,Zhang2013,Ebihara2019,Tsurutani2020}, including the May 2024 extreme storm during the solar cycle's maximum \cite{Hayakawa2025}, the underlying mechanism of the Solar-magnetosphere-ionosphere interaction remains poorly understood.

\begin{figure}[htbp]
\centering
\includegraphics[width=1.0\textwidth]{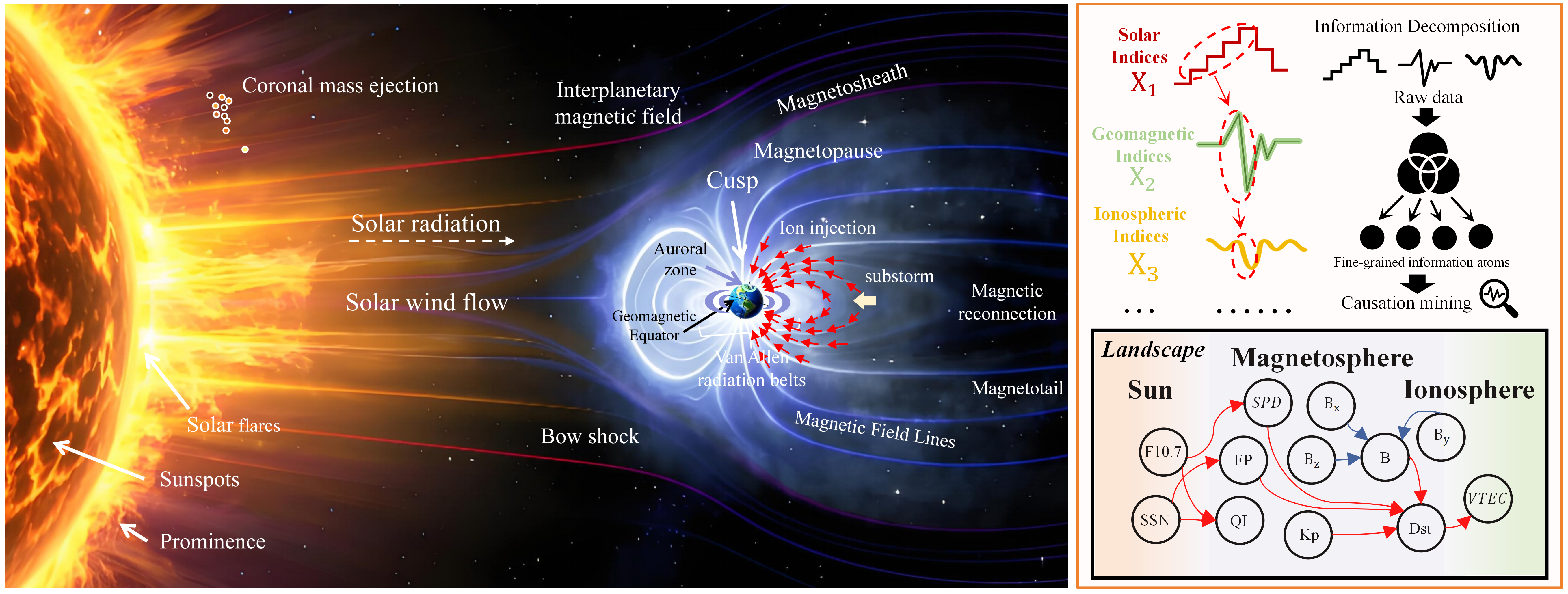}
\hspace{0.2cm}
\hfill
\caption{\textbf{The causal landscape for reconstructing the panoramic view of the Sun-Earth space weather during extreme solar storms.} The Solar-Magnetosphere-Ionosphere system involves particle and energy propagation of multiple physical processes, from which abundant multi-source observational data can be obtained. The potential mechanisms behind space weather remain to be explored from a data-driven perspective. The upper right part refers to information decomposition for causation mining for hidden interaction from spatio-temporal data. The lower right part shows an example of constructed causal landscape.} 
\label{fig:Fig1}
\end{figure}

Investigating physical mechanisms involves a systematic approach to understanding the underlying causes of observed phenomena in space. Nowadays, the advancement of observation technologies enables us to obtain more observational data \cite{Voosen2017} and data-driven science has emerged as the fourth paradigm alongside empirical, theoretical, and computational science. Without relying on predefined model assumptions for the underlying mechanisms, data-driven methods are uniquely suited to complex high-dimensional, nonlinear, and multiphysics coupling dynamic systems, like the solar space weather system \cite{Milan2017a}, where interactions are inherently non-trivial. Among the commonly used data-driven methods to infer causation \cite{Granger1969,Schreiber2000,Pearl2000,Liang2005}, the causality derived from information theory \cite{Shannon1948} has shown convincing evidence to unravel causal relationships \cite{runge2019,Runge2023,Stumpo2020}. However, the basic transfer entropy still requires an improvement in the disentangling of individual contributions from multiple sources. In recent years, the study of information decomposition and higher-order interactions has advanced significantly. Williams and Beer \cite{Williams2010} introduced a formal framework for decomposing information into unique, redundant and synergistic contributions, laying the foundation for modern partial information decomposition (PID)
\cite{Kolchinsky2022}. This advance has sparked significant interests in higher-order interactions \cite{Varley2023,Luppi2024} and highlights their potential applications in space science.  

\begin{figure}[htbp]
\sidesubfloat[]{\includegraphics[width=0.45\textwidth]{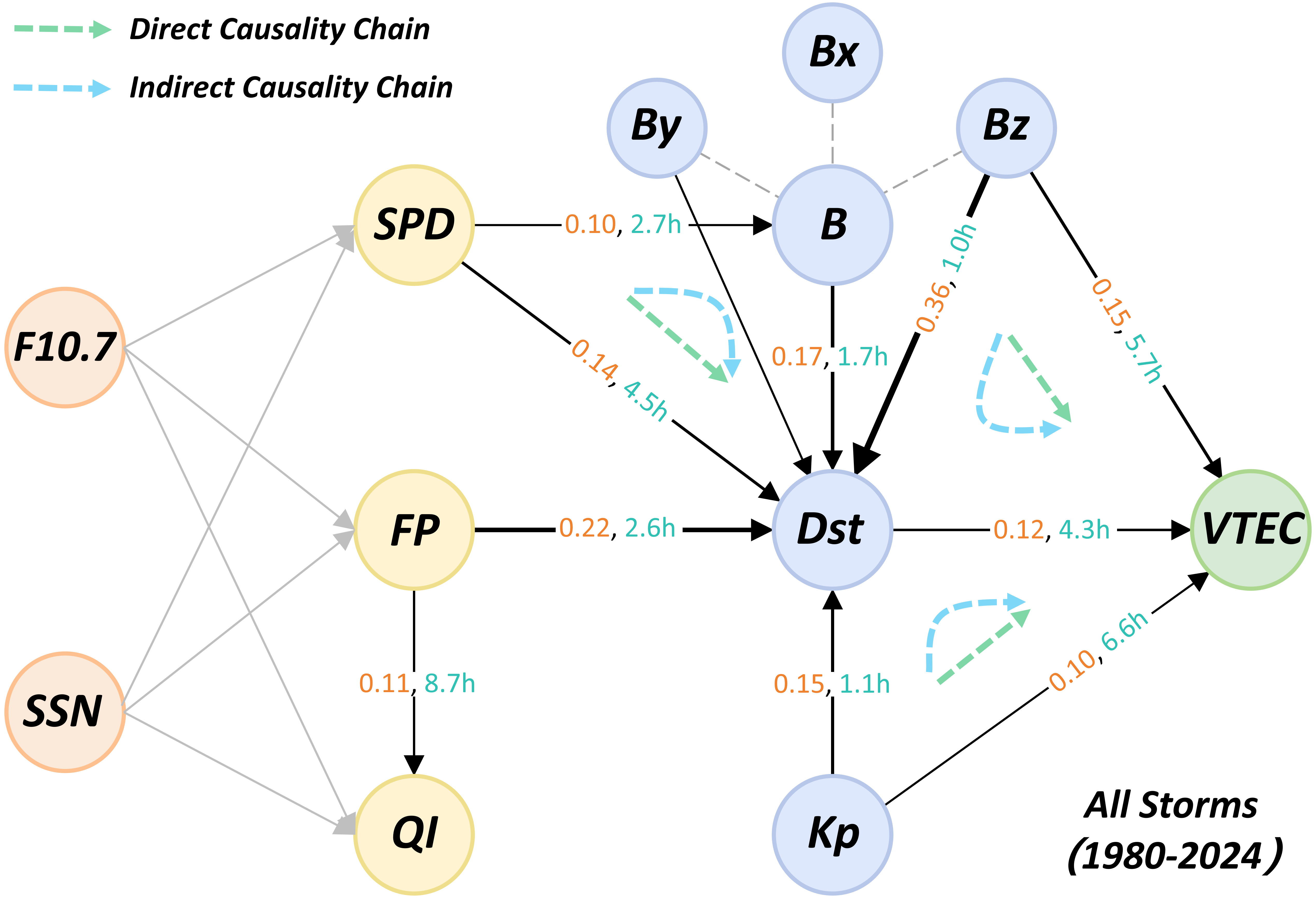}}
\sidesubfloat[]{\includegraphics[width=0.45\textwidth]{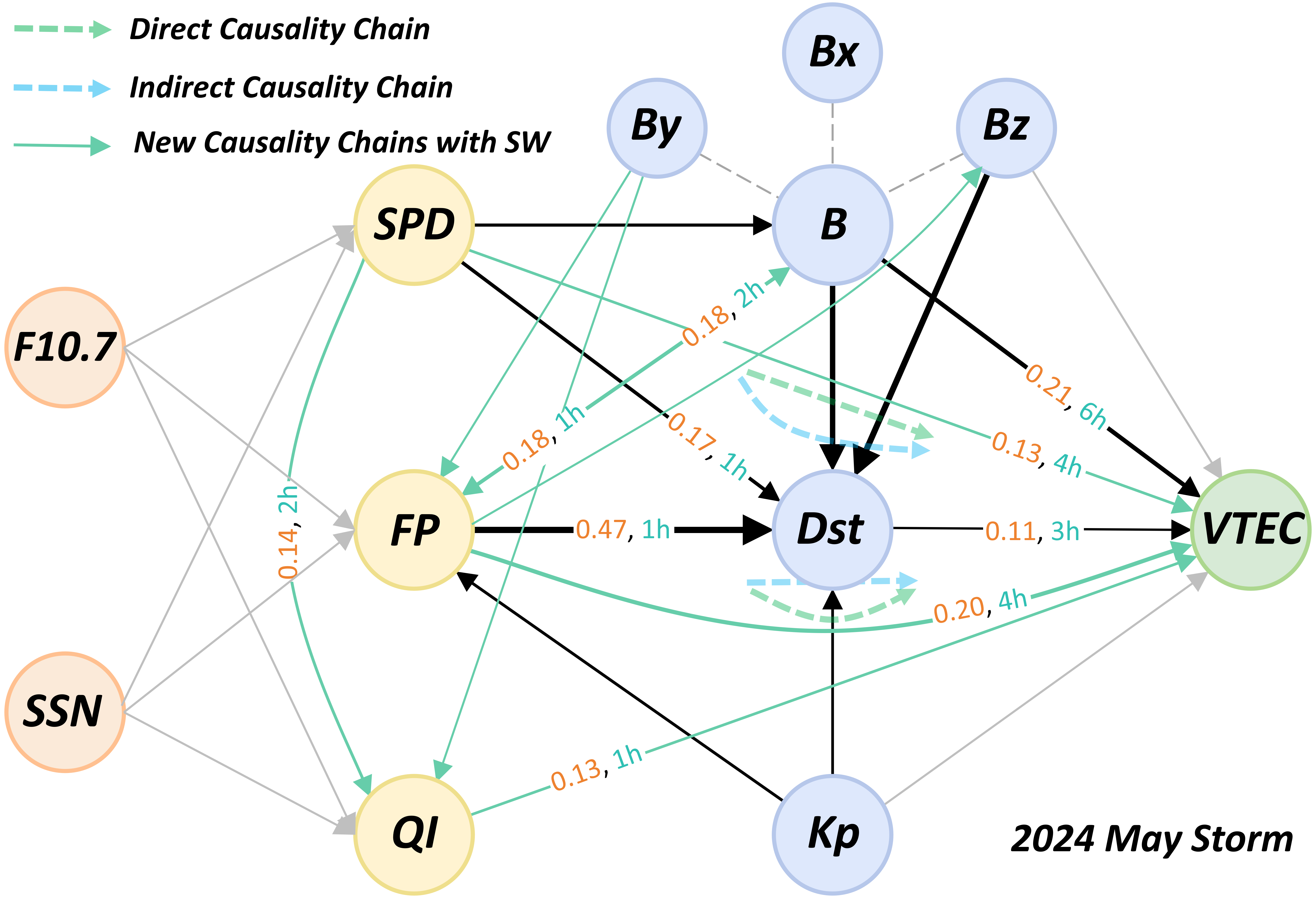}}\\
\vspace{0.5cm}  
\sidesubfloat[]{\hspace{0.2cm}\includegraphics[scale=0.8]{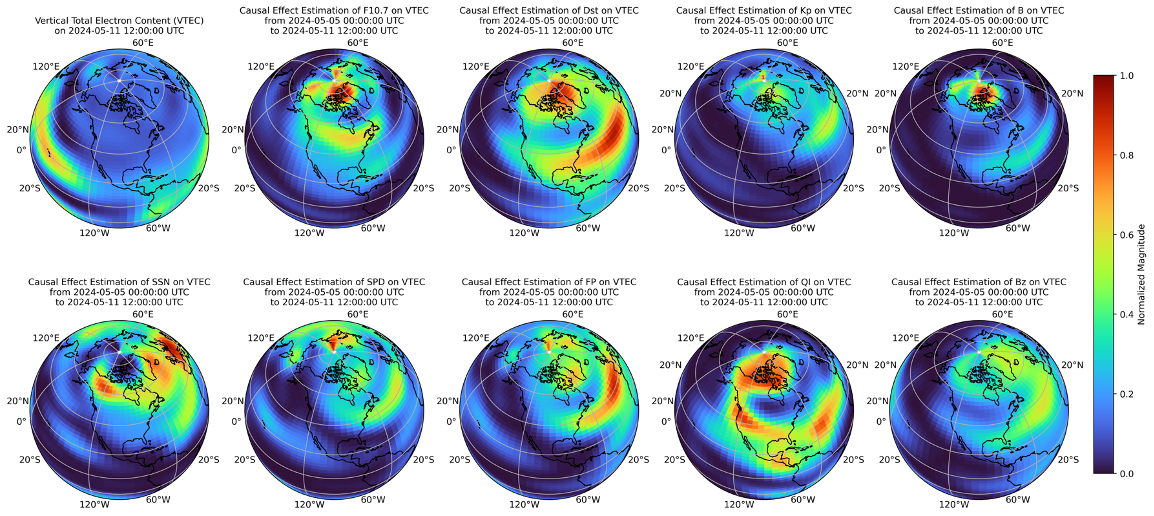}\label{fig:3a}}\\
\sidesubfloat[]
{\hspace{0.4cm}\includegraphics[scale=0.8]{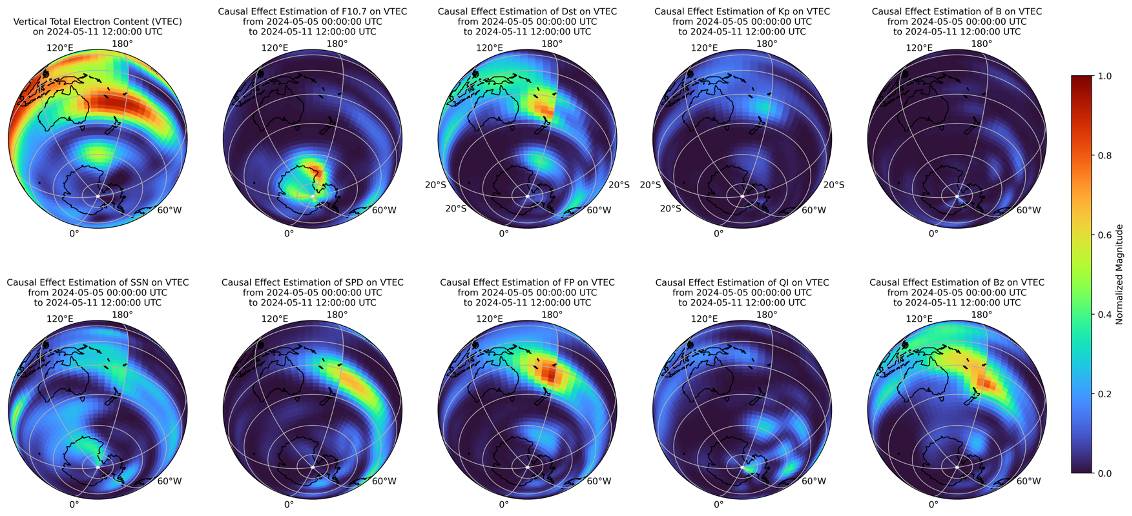}\label{fig:3b}}\\
\hfill
\captionsetup{font={small,stretch=1.25}}
\caption{\textbf{Constructed spatio-temporal causal graphs among space weather observable quantities.} 
(\textbf{A}) Directed causal graph for Statistical results from 273 geomagnetic storms (January 1, 1980 to October 14, 2024). 
(\textbf{B}) Directed causal graph for 2024 May Extreme Storm. 
(\textbf{C}) Global spatial causality map for the northern hemisphere. 
(\textbf{D}) Global spatial causality map for the southern hemisphere.}
\label{fig:Fig2}
\end{figure}

In this study, we pioneer an information-theoretic approach to uncover the intricate spatio-temporal causal landscapes within the solar-magnetosphere-ionosphere coupling system. A novel synergistic-unique-redundant decomposition method is introduced to extract latent causal patterns from heterogeneous and often low-quality solar storm data, as illustrated in Fig.\ref{fig:Fig1}. Based on the causality discovered, we develop SolarAurora, a machine learning model uniquely integrating causation into space weather prediction, unlike conventional atmospheric science models \cite{Pathak2022,Lam2023,Bi2023,Kochkov2024}. This SolarAurora model has been tested for predicting extreme solar storms in 2024, providing more generalizable causality-consistent forecasts with stronger result interpretability and enhancing our understanding of solar storm impacts.

\subsection*{\textbf{Synergy flux reveals spatio-temporal interaction landscapes}}

The May 2024 solar storms, occurring during Solar Cycle 25, unleashed a sequence of extreme solar flares, coronal mass ejections (CMEs), and caused geomagnetic disturbances from 10–13 May 2024. The associated  storm, reaching a provisional Dst index of -412 nT at 2 UT on 11 May, represents the Earth’s most intense event since March 1989 as the sixth-strongest recorded since 1957. This unprecedented activity triggered auroras visible at anomalously low latitudes, extending far into equatorial regions.

Causal diagrams in Fig.~\ref{fig:Fig2} illustrate the key cause-and-effect relationships governing space weather, from solar activity to its impacts on Earth's magnetosphere and ionosphere. The information transfer measured by synergy flux as shown in Fig.~\ref{fig:Fig2}A-B reveals well-defined hierarchical causal chains aligned with known physics in the solar-magnetosphere-ionosphere system. The causal pathway reflects the stepwise energy transfer and dynamic interactions governing geospace variability. Solar activity proxies (F10.7, SSN) correlate with solar wind parameters (SPD, FP, QI) observed in interplanetary space, which drives magnetospheric responses (Dst, Kp) via IMF Bz-dominated reconnection. These processes ultimately modulate the vertical ionospheric electron density (VTEC).

\begin{sloppypar}
In Fig.~\ref{fig:Fig2}A, the causal strengths and time delays, derived from statistical analysis of 273 geomagnetic storms, are shown along the causal pathways. The causal time delays exhibit high stability across different storms, revealing almost invariant lag characteristics in solar wind-magnetosphere coupling system. Notably, three causal loops with matched time delays are observed in all storms:  (1) an indirect/direct pathway from SPD$\rightarrow$B$\rightarrow$Dst/ SPD$\rightarrow$Dst; (2) an indirect/direct pathway from Bz$\rightarrow$Dst$\rightarrow$VTEC/ Bz$\rightarrow$VTEC ; (3) an indirect/direct pathway Kp$\rightarrow$Dst$\rightarrow$VTEC/ Kp$\rightarrow$VTEC. The temporal causal delay contributes to understanding the sequence of events, such as how a solar storm evolves into a geomagnetic storm and how different factors, like solar wind flow pressure or magnetic field orientation, affect the timing and severity of these storms.
\end{sloppypar}

Fig.~\ref{fig:Fig2}B shows causal diagrams during the May 2024 storm. By comparing with the Oct 2024 storm (Fig. S2 in supplementary materials), both storms exhibit analogous causal enhancements, including solar wind self-coupling (SPD$\rightarrow$Qi), intensified solar wind-magnetosphere pathways (SPD/FP$\rightarrow$B/Dst), and direct solar wind-ionosphere links (SPD/FP/Qi$\rightarrow$VTEC). Additionally, similar causal loops emerge, where the inclusion of SPD/FP$\rightarrow$VTEC introduces new loops such as SPD$\rightarrow$Dst$\rightarrow$VTEC and SPD$\rightarrow$VTEC, as well as FP$\rightarrow$Dst$\rightarrow$VTEC and FP$\rightarrow$VTEC. Both of these loops also follow the "matched time delays" pattern. 

The causal graph  (Fig.\ref{fig:Fig2}B) exhibits two key changes. 1) One feature is that solar wind exerts intensified direct control over magnetospheric-ionospheric systems via SPD$\rightarrow$Qi, FP$\rightarrow$B/Bz, and  FP/SPD/Qi$\rightarrow$VTEC. 2) Another feature is reciprocal solar wind-magnetosphere coupling FP$\leftrightarrow$B emerges via newly observed B$\rightarrow$FP (May) pathway.
Similar patterns are found, with the B$\rightarrow$Qi link appearing in the October 2024 storm and the FP$\leftrightarrow$B and Qi$\leftrightarrow$B couplings emerging in extreme geomagnetic storms ($Dst<-350$) (Fig. S1 in supplementary materials). Therefore, we speculate that solar wind-magnetosphere coupling patterns, such as FP$\leftrightarrow$B and Qi$\leftrightarrow$B, may be common to extreme geomagnetic storms.

Fig.\ref{fig:Fig2}C-D illustrate the spatial causality for interaction patterns for the May 2024 extreme storms. The global spatial causality of eleven parameters (F10.7, Dst, Kp, B, SSN, SPD, FP, QI, Bz, Bx, By) on global VTEC is calculated. During the storm phase\cite{Themens2024}\cite{Karan2024}, the spatial causality patterns (i.e., F10.7, SSN, Dst, Qi) shift predominantly to the Northern polar regions, a feature not captured by Gobal VTEC maps. For low latitude in the northern hemisphere, there appears high-causality region for observation variables (i.e., Dst, Qi, Fp, SPD and Bz). The evolution of the spatial causality map provides clearly high space weather impact geographical region.  

\begin{sloppypar}
Comparison of long-term causal connectivity (Fig. S1 in supplementary materials) with all storm results (Fig.~\ref{fig:Fig1}A) reveals marked changes. Over decadal timescales from two solar cycles (2002–2024), ionospheric variability shifts its dominant drivers from solar wind/magnetospheric perturbations to direct solar radiation, as F10.7/SSN$\rightarrow$VTEC causality strengthens while Bz/Dst/Kp$\rightarrow$VTEC weakens.
\end{sloppypar}

\subsection*{\textbf{Discovery and Attribution of Auroral Spatial Causality Patterns}}
\begin{figure}[htbp]
\centering
\sidesubfloat[]{
    \includegraphics[scale=0.53]{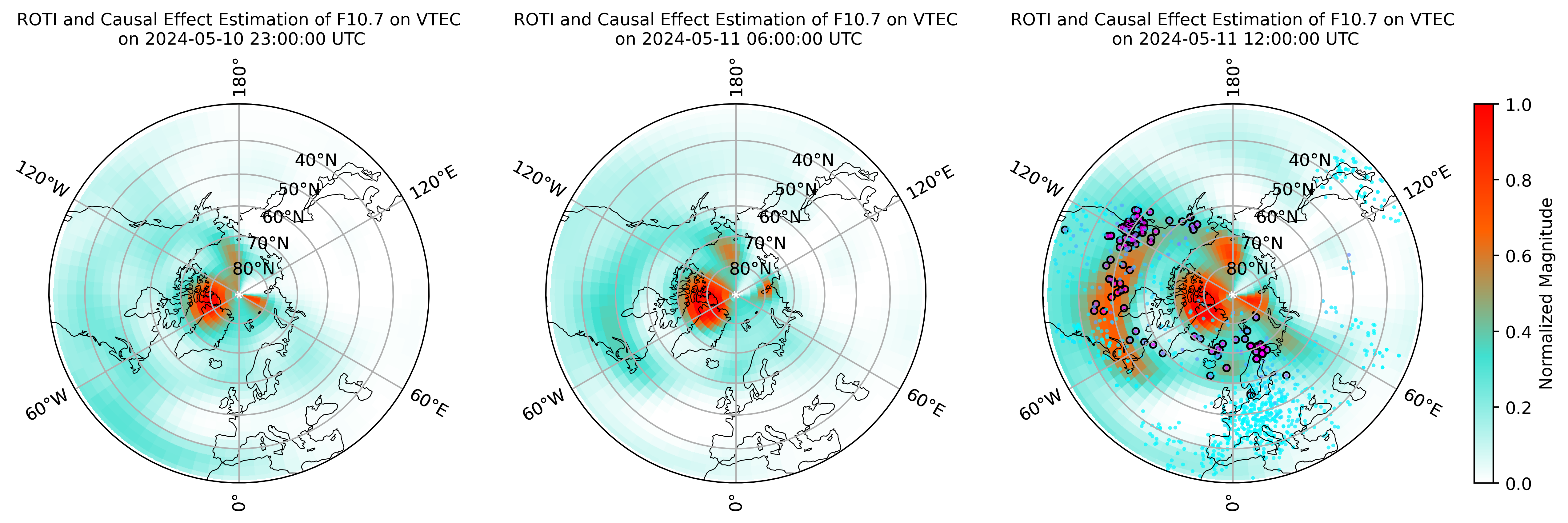}\label{fig:3c}}\\
\sidesubfloat[]{
    \includegraphics[scale=0.5]{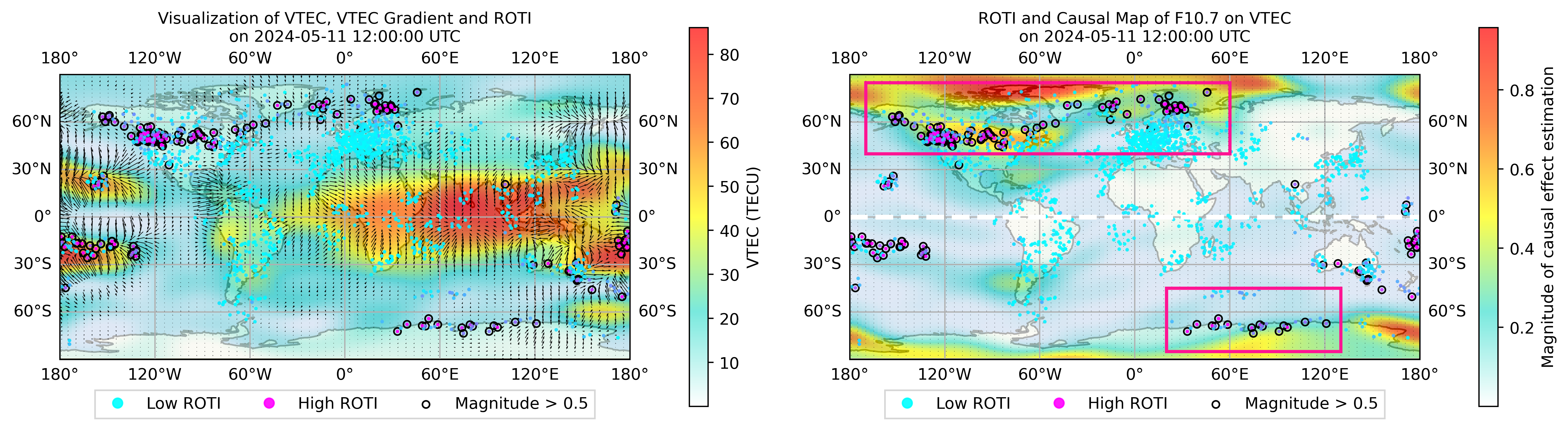}\label{fig:3d}}
\caption{{\textbf{Spatio-temporal causal patterns of the solar-space environment during the May 2024 extreme solar storm.} (\textbf{A}) A temporal evolution collection of auroral spatial causality map from F10.7 to VTEC in the Northern hemisphere. (\textbf{B}) GNSS ROTI map with VTEC and causality maxima moments in geographical coordinates. For visualization convenience, the causal influence is Min - Max normalized.}}
\label{fig:Fig3}
\end{figure}

Information flow analysis reveals causal coupling mechanisms in physical processes. The absence of a direct causal link (edge) between two variables does not necessarily imply the lack of a causal relationship, but may reflect measurement limitations (e.g., inadequate spatio-temporal resolution and measurements confined to certain areas). The spatial causality map provides global space weather dynamics, revealing high-order interactions often obscured by conventional methods.

Fig.\ref{fig:Fig3} illustrates the spatial causality interaction patterns for the May 2024 extreme storms.  During the storm phase, the spatial causality patterns (i.e., from F10.7 and SSN) shift predominantly to the Arctic, a feature not captured by VTEC maps. Rapid evolution of auroral spatial causality patterns between F10.7 and VTEC is observed (Fig.~\ref{fig:Fig3}A), with intensified activity in the northern auroral region at UT 12 on May 11. The coverage and shape of such auroral spatial causality patterns aligns with previous auroral oval observations \cite{Zhang2021}\cite{Qing-He2013}. 

Similar spatial causality dynamics are noted during the Oct 11, 2024 storm (Dst = $\sim -341 \,\rm{nT}$) (Fig. S3 in supplementary materials), suggesting a recurrent feature of extreme geomagnetic storms. The reason for strong auroral spatial causality may be explained as during extreme storms, solar wind particles penetrate to lower altitudes, which may be reflected by the DMSP observations \cite{Strom2005} and enhanced Pedersen current density, and Hall current density at $125\, \rm{km}$ as modeled by SAMI3 \cite{Huba2008} (Fig. S4 in supplementary materials). 

Global storm impacts GNSS signals in high-auroral-causality regions, which exhibit elevated ROTI (ionospheric irregularity index) in Figs.\ref{fig:Fig3}A–B. The northern auroral spatial causality is consistent with the power grid stress experienced in North America similarly as on March 13, 1989 (Dst  $\sim -597 \,\rm{nT}$) causing 9-hour blackout of Hydro-Québec’s electricity transmission due to geomagnetically induced currents (GICs). Therefore, these auroral spatial causality patterns in Figs.\ref{fig:Fig3}A can serve as a new indicator for evaluating the storm's impact region. The underlying physical mechanisms can be further validated through the upcoming Solar Wind Magnetosphere Ionosphere Link Explorer (SMILE) mission \cite{Wang2025} and future storms during the next solar cycle. 

\subsection*{\textbf{Quantitative Evaluation of Causality Learning}}
\label{sec2}

Here, we assess the information-theoretic method on existing machine learning models via multiple-input single-output randomized experiments, where input channels are selected determined by the synergy flux. Randomized experiments (Fig. S5 in supplementary materials) reveal that F10.7 prediction roughly degrades with added input channels, whereas Dst forecasting initially improves but eventually declines as channels are added in descending causal order. Using 24-hour RMSE as a metric, performance decline is attributed to redundant noise from weakly causal variables. These results validate the information-theoretic method’s potential to enhance ML model interpretability and performance by prioritizing causal relevance.

\begin{figure}[H]
\centering
\includegraphics[width=1.0\linewidth]{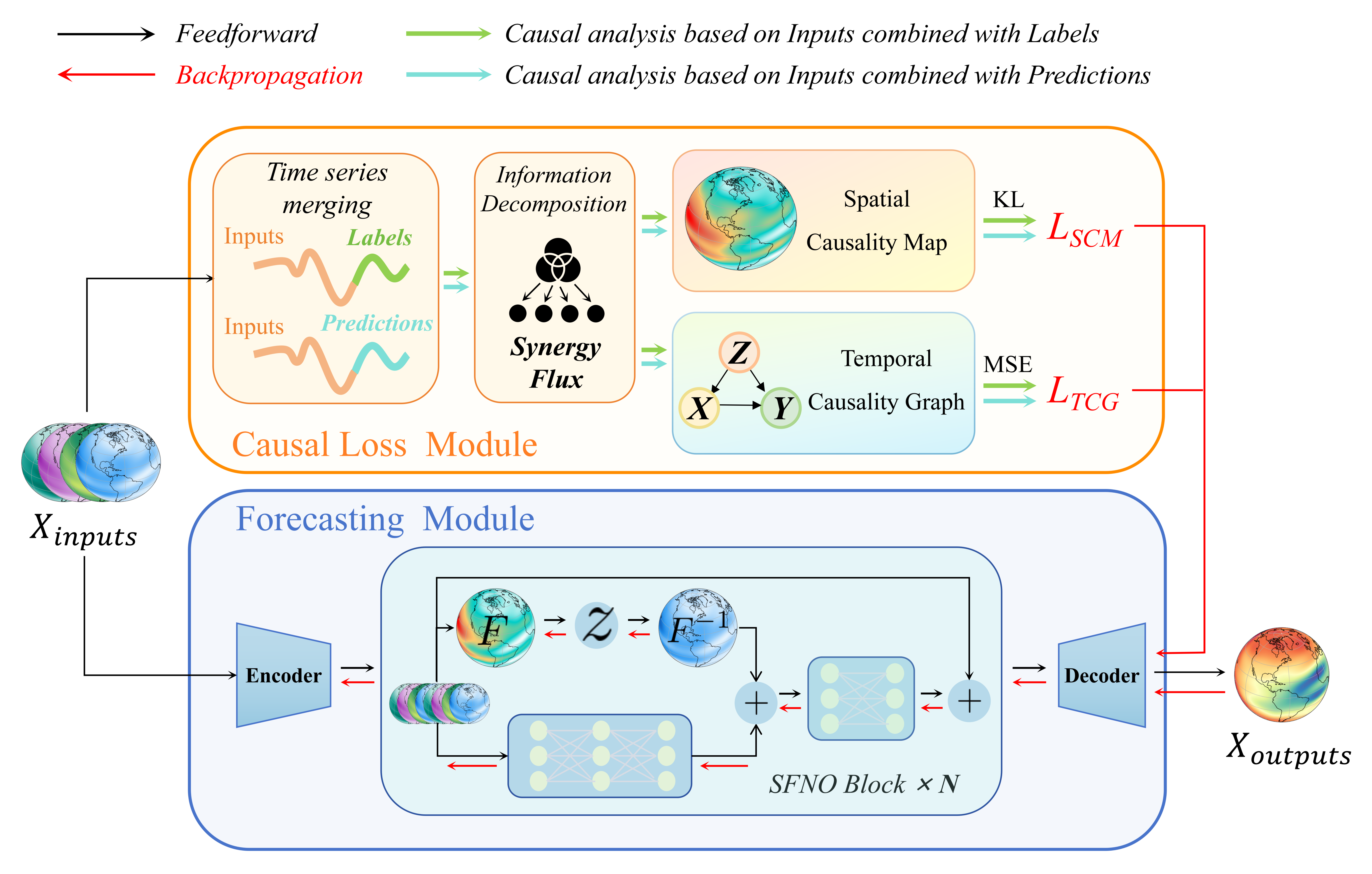} 
\caption{\textbf{A schematic diagram of SolarAurora Model for space weather forecasting.} The model integrates a forecasting module with a causal loss module. The Causal Module (upper panel) supervises the model by enforcing spatio-temporal causal consistency, using synergy flux derived from information decomposition to construct spatial causal maps (SCM) and temporal causal graphs (TCG). The Forecasting Module (lower panel), as a surrogate model of real physical processes, predicts future space weather states (F10.7, Kp, Dst, VTEC, etc.) through an encoder–SFNO blocks–decoder structure.}
\label{fig:Fig4}
\end{figure}

Information atoms possess the remarkable ability to characterize cross-channel information interactions within black-box models. Inspired by this concept, we developed SolarAurora, a causality-informed machine learning model designed to enhance both the accuracy and interpretability of space weather prediction.

Figure~\ref{fig:Fig4} illustrates the schematic of SolarAurora. The model inputs include diverse observables from the solar–terrestrial environment, varying in spatial resolution and temporal sampling. The prediction backbone encodes the latent physical dynamics using a deep neural architecture composed of an encoder, a stack of Spectral-Fourier Neural Operator (SFNO) layers, and a decoder. This structure enables effective representation learning over the spherical manifold and efficient feature extraction in both time domain and frequency domain. In addition, SolarAurora is trained with a hybrid loss function that incorporates causal-consistency regularization. By leveraging the synergy flux derived from information decomposition, the model enforces causal coherence across both spatial and temporal dimensions during training. SolarAurora supports fully end-to-end training without requiring external interventions or handcrafted features.

\subsection*{Causality-informed extreme space weather prediction}

\begin{figure}[H]
\centering
\begin{minipage}[t]{0.715\textwidth}
  \renewcommand{\thesubfigure}{A}
  \sidesubfloat[]{\includegraphics[width=\textwidth]{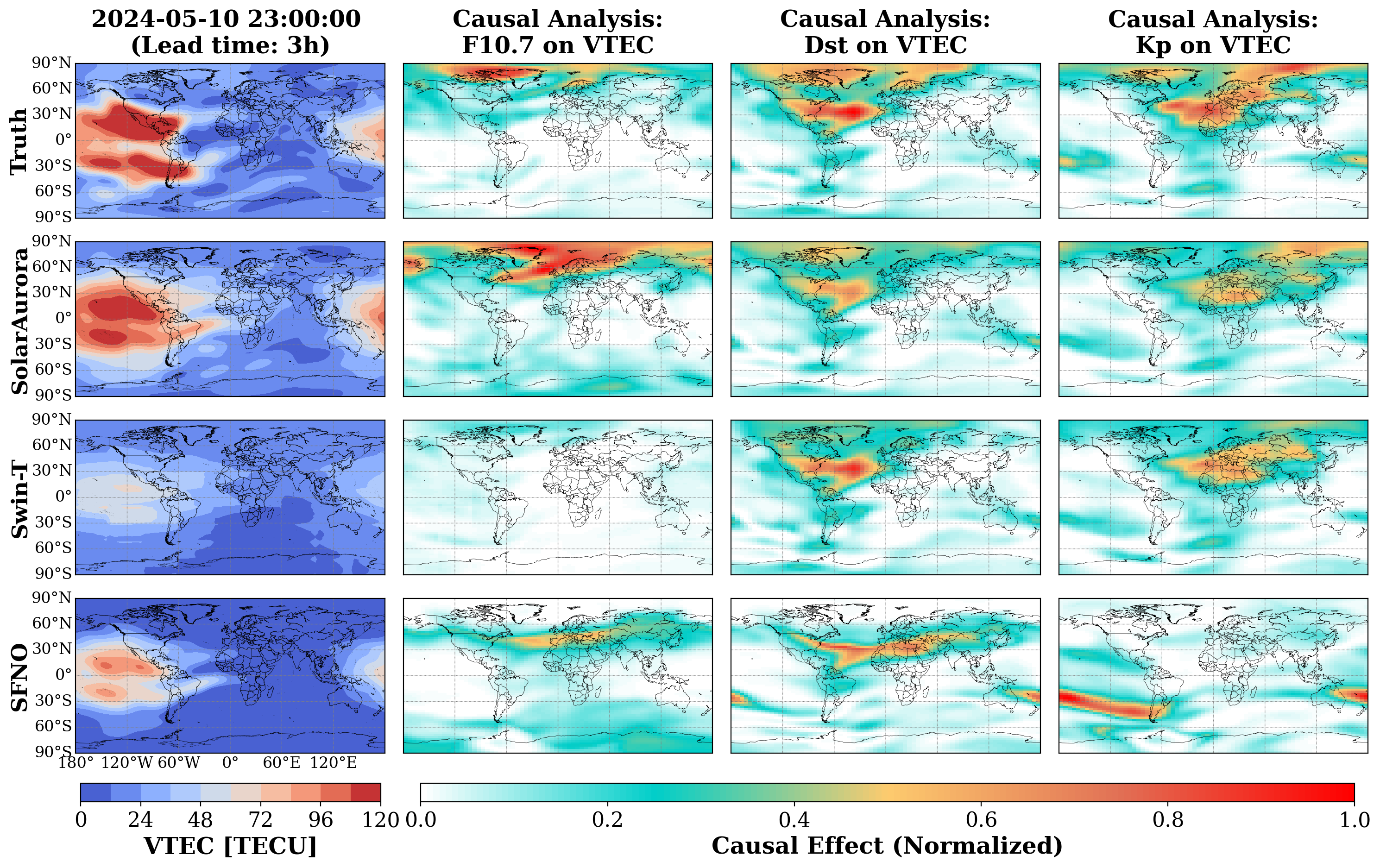}}\par
  \renewcommand{\thesubfigure}{C}
  \sidesubfloat[]{\includegraphics[width=\textwidth]{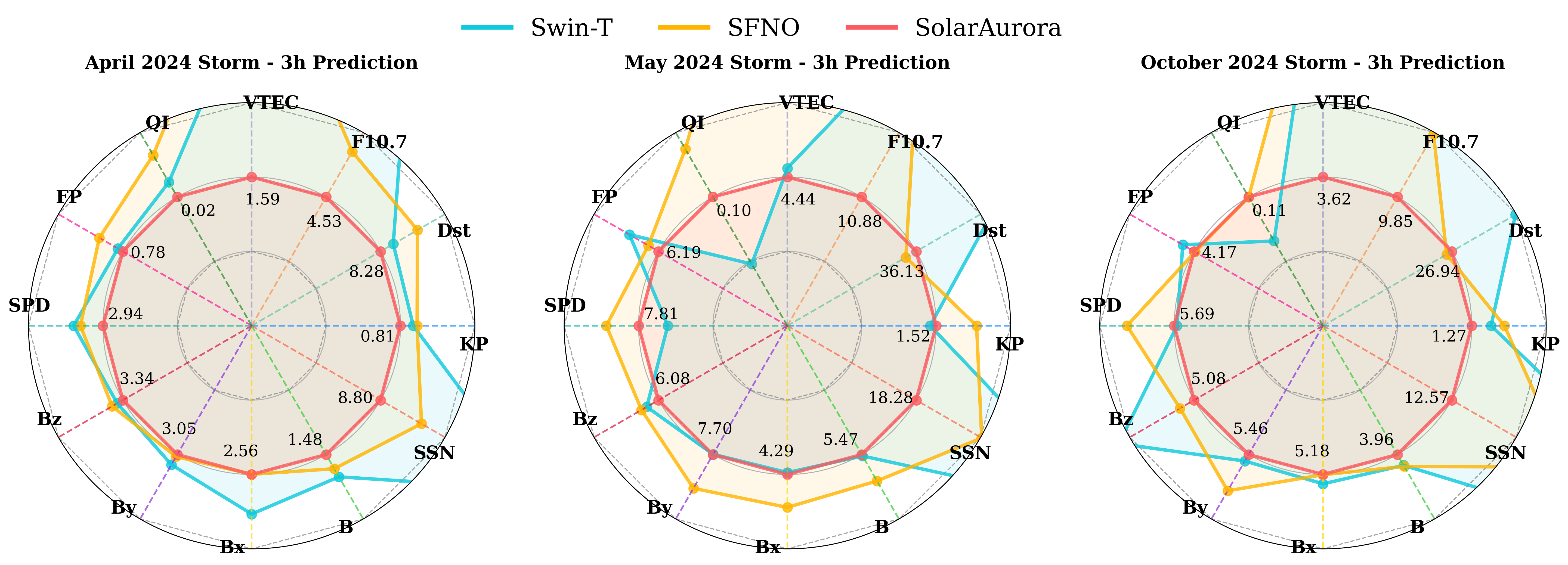}}
\end{minipage}
\hfill
\begin{minipage}[t]{0.26\textwidth}
  \renewcommand{\thesubfigure}{B}
  \sidesubfloat[]{%
    \begin{minipage}[t]{\textwidth}
      \vspace*{-3.5mm} 
      \includegraphics[width=\textwidth]{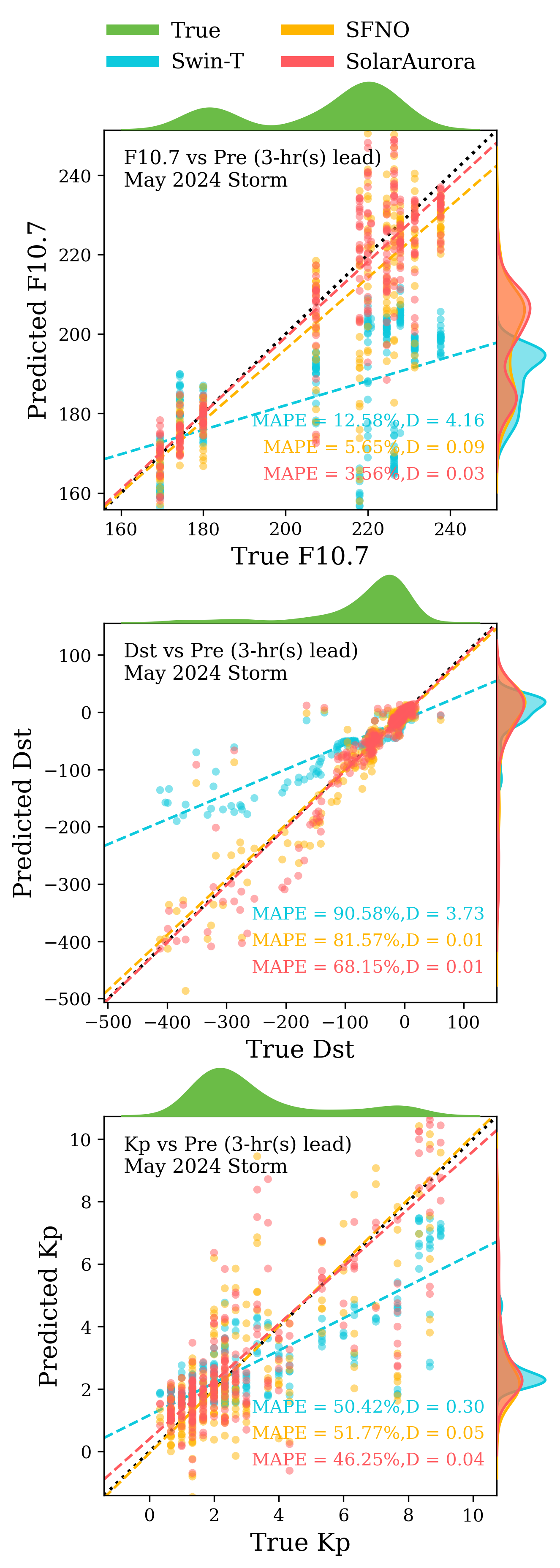}
    \end{minipage}
  }
\end{minipage}
\caption{\textbf{Statistical quantitative evaluation of SolarAurora during the 2024 solar extreme storms.} (\textbf{A}) comparison of the 3-hour prediction of ionospheric VTEC and the causality map during the May 2024 storm is conducted at 23:00:00 UTC on May 10, 2024. 
(\textbf{B}) comparison of the 3-hour ahead prediction of F10.7, Dst, and Kp for SolarAurora during the 2024 May extreme storm. (\textbf{C})comparison of 12-variables prediction using RMSE for three strong storms in Apr 12-23, May 04-15, and Oct 04-14, 2024.}
\label{fig:Fig5}
\end{figure}

To evaluate SolarAurora, we collect typical 12 observational variables of solar storm dataset for proof-in-concept demonstration as explained in Supplementary Table \ref{table1}. The horizontal resolution for ionospheric vertical total electron content (VTEC) map is  $2.5^{\circ}$ in latitude and $5^{\circ}$ in longitude. The training dataset is from Oct 19, 2014 to March 29, 2024. The prediction performance of SolarAurora is evaluated with Spherical Fourier Neural Operators (SFNO) \cite{Bonev2023} and Swin-Transformer \cite{Vaswani2017,Liu2021} for benchmark.

Fig.\ref{fig:Fig5} presents the statistical quantitative evaluation of SolarAurora during the 2024 geomagnetic storms. Over the extreme storm in May 2024, compared with SFNO and Swin-transformer(Swin-T), SolarAurora shows the best 3-hour prediction performance for the VTEC map and spatial causality map at the geomagnetic storm time UT 23:00 on May 10 in Fig.\ref{fig:Fig5}a, as well as for F10.7, Dst and Kp in Fig.\ref{fig:Fig5}b. SolarAurora reduces MAPE by $9.02\,\%$, $22.43\,\%$, and $4.17\,\%$ compared to Swin-Transformer, and by $2.09\,\%$, $13.42\,\%$, and $5.52\,\%$ compared to SFNO, for F10.7, Dst, and Kp, respectively, with the lowest KL divergence among all models. For the prediction of all 12 variables during the 2024 (April, May, October) geomagnetic storms in Fig.\ref{fig:Fig5}c, the radar charts of the relative root mean square error (RMSE) demonstrate that our SolarAurora achieves balanced and stable prediction. 

\subsection*{Discussion}
\label{sec4} 
The causal propagation chain of space weather interactions during solar storms has been systematically investigated, with particular emphasis on the extreme solar storm of May 2024. Using a synergistic-unique-redundant information decomposition framework, we successfully captured rich spatio-temporal causal structures underlying the complex physical interactions within the solar-magnetosphere-ionosphere coupling system. To the best of our knowledge, this work is the first to reveal such a spatiotemporal causal graph for the coupling system during an extreme event, based on an information-theoretic approach. These insights enabled the development of SolarAurora, a novel prediction model that integrates causality and significantly enhances the prediction performance for space weather phenomena, demonstrating the utility of causation mining in space science.

Our study presents compelling evidence for stable spatiotemporal causality patterns in the solar-terrestrial environment, derived from historical space weather datasets. During the May 2024 extreme event, we identified two new dynamical behaviors: first, intensified direct control of the solar wind over the magnetospheric and ionospheric systems (e.g., SPD/FP$\rightarrow$B/Dst, SPD/FP/Qi$\rightarrow$VTEC); second, reciprocal coupling between solar wind and interplanetary magnetic field (e.g., FP$\leftrightarrow$B). Associated causal time delays, such as 4.5 hours for SPD$\rightarrow$ Dst and 4 to 6 hours for magnetosphere-to-ionosphere links, remained highly consistent in 273 geomagnetic storms. These temporal features suggest their potential utility for early warning. Relative to short-term storm analyses, long-term causal analysis (Fig. S1 in the supplementary materials) shows that ionospheric drivers have shifted from solar wind/magnetospheric perturbations to direct solar radiation, as F10.7/SSN$\leftrightarrow$VTEC causality strengthens while Bz/Dst/Kp$\leftrightarrow$VTEC weakens.

In addition, a spatially resolved causal map of auroral activity captured during the extreme event highlighted the strong coupling between solar and ionospheric systems. In particular, clear causal connections are detected in the northern polar auroral region—relationships that would have remained hidden if only single-source measurements such as GNSS-based VTEC were used. This highlights that intervariable causality, particularly synergistic information, may expose latent physical interactions that are often concealed in noisy or low-resolution datasets, unlike conventional correlation-based approaches (Fig. S2 in supplementary materials). This also means that the work provides a brand new criterion for the field.

In contrast to conventional models, SolarAurora embeds the capacity to perceive causal knowledge into machine learning frameworks, functioning as a spatio-temporal attention mechanism that improves both predictive accuracy and interpretability. By incorporating both temporal and spatial causality, it improves understanding of how solar storms evolve into geomagnetic disturbances and their regional impacts. It achieves superior performance in forecasting the peak distribution of VTEC during the May 2024 storm, and showed consistent gains across multiple key parameters. For example, in the case of Dst, SolarAurora outperformed the baseline SFNO model by 13.42\% in mean absolute percentage error (MAPE), and consistent improvements were observed across most of the 12 evaluated variables during geomagnetic storms.

However, data-driven methods fundamentally depend on data quality and the precision of causality estimation. Accurately quantifying information synergy and redundancy remains a technical challenge. In this study, we adopted approximation strategies to overcome the limitations imposed by sparse or noisy data and computational constraints. Moreover, causal time delays, which revealed clear temporal dynamics, have not yet been explicitly incorporated into the model architecture. Future research will aim to develop more efficient and scalable causality-aware frameworks to extract hidden dynamics from large-scale space weather data, further advancing the field of data-driven geospace forecasting.


\clearpage 

%
\bibliography{science_template} 
\bibliographystyle{sciencemag}

%
%
%
%
%
%


\section*{Acknowledgments}
We are grateful to Prof. X. S. Liang for inspiring discussion on causality theory. We appreciate the initial attempts of the prediction model developed by S. D. Jiang, Z. H. Shi, S. Zhou and data preparation by S. Zhou, J. Y. Ma, Y. Sui. The computing is performed using CFFF(Computing for the Future of Fudan University) platform \textcolor{blue}{https://cfff.fudan.edu.cn/.}
\paragraph*{Funding:}
This work was funded by the National Key Research and Development Program of China \# 2021YFA0717300 (H.Y.Fu).
\paragraph*{Author contributions:}
Conceptualization: Haiyang Fu; Methodology: Xinan Dai; Data collection: Zichong Yan, Zitong Wang; Experiment: Zichong Yan, Xinan Dai, and Zitong Wang; Funding acquisition: Haiyang Fu; Supervision: Haiyang Fu; Writing–original draft: Xinan Dai and Haiyang Fu; Writing–review \& editing: Haiyang Fu, Feng Xu, Chi Wang, Ya-Qiu Jin.
\paragraph*{Competing interests:}
There are no competing interests to declare.
\paragraph*{Data and materials availability:}
The dataset analyzed in the current study and the statistical script are available in \textcolor{blue}{https://github.com/SolarAuroraRelease/SolarAurora.}



\end{document}